\documentclass[a4paper,11pt]{article}
\usepackage[position=t,singlelinecheck=off]{subfig}
\usepackage{amssymb,amsmath}
\usepackage{graphics,graphicx,float}
 \usepackage{mathabx}
\usepackage{epstopdf}
\usepackage[affil-it]{authblk}

\def\beq{\begin{equation}}
\def\eeq{\end{equation}}
\def\bea{\begin{eqnarray}}
\def\eea{\end{eqnarray}}
\def\nn{\nonumber}

\def\lo{\left(}
\def\rc{\right)}

\makeatletter
 
  \def\@cite#1#2{${\mbox{#1\if@tempswa , #2\fi}}$}
\makeatother

\title{\LARGE{\sf{Evolution of nonclassicality of the quasi-Bell states for a strongly coupled qubit-oscillator system}}}
\author{R. Chakrabarti and B. Virgin Jenisha}
\affil{Department of Theoretical Physics, University of Madras, Chennai-600 025, India}
\date{}
\begin{document}
\maketitle
\begin{abstract}
 Starting with the quasi-Bell states of the qubit-oscillator system, we obtain  time evolution of the density matrix under the adiabatic approximation. The composite density matrix leads to, via partial tracing of the qubit degree of freedom, the  reduced density matrix of the oscillator
  that is utilized to obtain the quasi-probability distributions such as
Glauber-Sudarshan $P$ function, Wigner $W$ function and Husimi $Q$ function. 
The negativity of the Wigner function acts as
 a measure of the  nonclassicality of the state.
The negativity becomes particularly relevant in understanding a comparison between the Wigner entropy  with the Wehrl entropy, which are based on the $W$ function and $Q$ function, respectively.  
\end{abstract}

\section{Introduction}
A two-level system (qubit) that interacts with a radiation field represented by a single oscillator mode
 is one of the important models in quantum optics. 
This model has been studied extensively 
under rotating wave approximation (RWA)\,[\cite{JC1963}] that holds good for weak coupling regime.
Following the recent developments in experiments [\cite{ABS2002}-\cite{ND2010}], it is of interest to study the system with higher 
qubit-oscillator coupling.
In the (ultra) strong coupling domain and for low qubit frequency where the oscillator frequency sets the scale, the system has been investigated using adiabatic 
approximation [\cite{IGMS2005,AN2010}] that utilizes the separation of the time scales of the high oscillator frequency  and the low (renormalized) frequency of the qubit.

\par

To study the dynamics of the quantum state, we employ various quasi-probability distributions on the phase space as preferred tools.
For instance, the Wigner function [\cite{W1932}] gives a connection between the classical and quantum dynamics.
Unlike the classical (true) probability distributions, the Wigner function $W$ may assume negative values. This negativity [\cite{KZ2004}] serves  as an indicator 
of the nonclassicality of the quantum state. On the other hand, entropy is a key concept for
the quantum systems that provides a framework for studying the loss of information.
The $W$ distribution has also been recently employed [\cite{MAN2000, SKD2012}] for developing a  new measure of quantum entropy. This is found to 
have close correspondence with the negativity parameter.

\par

The present manuscript is organized as follows: The initial quasi-Bell composite state and the  corresponding reduced density matrix of the oscillator are considered
under adiabatic approximation in Sec.II. This density matrix is utilized in Sec.III to derive the Glauber-Sudarshan $P$ distribution [\cite{G1963,S1963}]. 
In Sec.IV we evaluate the Wigner $W$ function [\cite{W1932}] which is used to study the nonclassicality and decoherence of the system. 
The relevant Husimi $Q$ distribution [\cite{H1940}] admits, in the weak coupling regime, a closed from evaluation
\textit{\`{a} la} the procedure employed in [\cite{IGMS2005}]. Comparison of the time evolution characteristics of the Wherl entropy [\cite{W1978}] based on the $Q$
distribution, and that of the Wigner entropy [\cite{MAN2000,SKD2012}], which is based on the $W$
distribution, is discussed in the context of negativity of the associated quantum state.
 \section{Reduced density matrix of the oscillator}
We study a coupled qubit-oscillator system with the  Hamiltonian [\cite{IGMS2005},  \cite{AN2010}] that reads 
in natural units $(\hbar=1)$ as follows:
\beq 
 H = -\frac{\Delta}{2} \sigma_x - \frac{\epsilon}{2} \sigma_z +
 \omega a^{\dagger}\, a + \lambda  \sigma_z \,(a^{\dagger} + a),
 \label{Hamiltonian}
\eeq
where the harmonic oscillator with a frequency $\omega$ is described by the raising and lowering operators 
$(a^{\dagger}, \,a| \hat{n} \equiv a^{\dagger} a)$. The qubit characterized by an energy splitting $\Delta$ as well as an external 
static bias $\epsilon $ is expressed via the spin variables $(\sigma_x, \,\sigma_z)$. The qubit-oscillator coupling strength is denoted by $\lambda$.  The Fock states 
$\{\hat{n} |n\rangle = n |n\rangle,\,n = 0, 1,\ldots;\;a \,|n\rangle = \sqrt{n}\,|n - 1\rangle, 
a^{\dagger}\, |n\rangle = \sqrt{n + 1}\,|n + 1\rangle\}$ provide the basis for the oscillator, whereas the eigenstates  $\sigma_z |\pm 1\rangle = \pm \,|\pm 1\rangle$ span the space of the qubit.  The Hamiltonian (\ref{Hamiltonian}) is not known to be exactly solvable.
In the present work we follow the adiabatic approximation [\cite{IGMS2005},  \cite{AN2010}] that hinges on the separation of the time scales governed by the high oscillator frequency and the (renormalized) low qubit frequency: $\omega \gg \Delta$. 

\par

We consider the following quasi-Bell initial states of the coupled system
\beq
|\psi(0)\rangle^{(\pm)} = \dfrac{1}{\sqrt{2}}\left(|1,\alpha \rangle \pm |-1, -\alpha \rangle\right),\qquad
|\alpha \rangle = \mathrm{D}(\alpha) |0 \rangle, \qquad
 \mathrm{D}(\alpha)= \exp(\alpha a^{\dagger}-\alpha^*a),
\label{quasi_Bell}
\eeq
where the coherent state  $\{|\alpha \rangle \; \forall \alpha = \mathrm{Re}(\alpha) + i \mathrm{Im}(\alpha) \in \mathbb{C}\}$ for the oscillator degree of freedom is realized by the action of the 
displacement operator on the vacuum. Under adiabatic approximation [\cite{IGMS2005}], the evolution of the initial state (\ref{quasi_Bell})
yields the  bipartite density matrix: 
\beq
\rho^{(\pm)}(t) = |\psi(t)\rangle^{(\pm)}\langle \psi(t)|.
\label{density_composite}
\eeq
The partial tracing, say, over the qubit degree of freedom produces the reduced density matrix of the oscillator: 
\beq
\rho_{\cal O}^{(\pm)}(t) \equiv \hbox {Tr}_{\cal Q}\,|\psi(t)\rangle^{(\pm)}\langle \psi(t)|
\label{o_density}
\eeq
that may be expressed [\cite{R2013}], in terms of the displaced number states $|n_{\pm} \rangle=\mathrm{D}^{\dagger}\left(\pm \lambda/\omega \right) | n \rangle$ as follows:
\bea
\rho_{\cal O}^{(\pm)}(t)&=&\dfrac{1}{2} \exp(-|\alpha_{+}|^2)\sum_{n,m=0}^{\infty} 
  \dfrac{(\alpha_{+})^{n} \,  (\alpha_{+}^*)^{m}}{\sqrt{n!m!}} 
 \Big ({\mathfrak{C}_n^{\pm}(t)\,\mathfrak{C}_m^{\pm}}(t)^* |n_+ \rangle \langle m_+ |\nn \\
&& + (-1)^{n+m}
{\mathfrak{C}_n^{\mp}}(t)^* \,\mathfrak{C}_m^{\mp}(t) |n_- \rangle \langle m_- | \Big) \exp \big (-i (n - m)\omega t \big).
\label{o_density_matrix}
\eea
The coefficients in (\ref{o_density_matrix}) are given by
\beq
\mathfrak{C}_n^{\pm}(t)=  \mathfrak{A}_n^{\mp} \exp (i \chi_n t)+\mathfrak{B}_n^{\pm} \exp (-i \chi_n t),
 \quad \mathfrak{A}_n^{\pm}= \dfrac{\chi_n+\tilde{\epsilon} \pm (-1)^n \delta_n}{2 \chi_n},
\quad \mathfrak{B}_n^{\pm}= \dfrac{\chi_n-\tilde{\epsilon} \pm (-1)^n \delta_n}{2 \chi_n}, \nonumber
\label{o_density_ab}
\eeq
where, $\chi_n=\sqrt{\delta_n^2+\tilde{\epsilon}^2}$,\,
$\delta_n = -\dfrac{\widetilde{\Delta}}{2}  L_{n}(x), \; x = \left( 2\lambda/ \omega\right)^2$ ,\;
$\widetilde{\Delta} = \Delta \exp \lo - \frac{x}{2}\rc,$ 
$\tilde{\epsilon}=\dfrac{\epsilon}{2}$
 and 
$\alpha_{+}=\alpha + \lambda /\omega$.
The  reduced density matrix (\ref{o_density_matrix}) of the oscillator may be utilized to calculate the quasi-probability distributions.
\section{Phase space distributions}
\subsection{Glauber-Sudarshan $P$ function}
The well-known Glauber-Sudarshan $P$ function [\cite{G1963,S1963}] admits  a diagonal 
representation of the oscillator density matrix in the coherent state basis:
\bea 
\rho= \int P(\beta) | \beta \rangle \langle \beta | d^2 \beta.
\label{rho-P}
\eea
 For an arbitrary quantum state the relation (\ref{rho-P}) may be inverted and the $P$ function is uniquely expressed as
\bea
P(\beta)=\dfrac{e^{|\beta|^2}}{\pi^2}\int \langle -\gamma | \rho|\gamma \rangle e^{|\gamma|^2} e^{-\gamma \beta ^*+\gamma^* \beta} d^2 \gamma.
\eea
For our choice of the reduced density matrix (\ref{o_density_matrix}) the $P$ function may be derived readily as follows:
\bea
P^{(\pm)}(\beta) &=& \dfrac{1}{2} \exp(-|\alpha_{+}|^2) \sum_{n,m=0}^{\infty} \dfrac{\alpha_{+}^n \alpha_{+}^*{}^m}{n!m!}
\Big(\frac{\partial}{\partial \beta}\Big )^n \Big(\frac{\partial}{\partial \beta^*}\Big )^m \times \nn \\
&& \times \Big((-1)^{n+m} \mathfrak{C}_{n}^{\pm}(t) \mathfrak{C}_{m}^{\pm}(t)^* e^{|\beta_{+}|^2} \delta^{(2)}(\beta_{+}) 
+\mathfrak{C}_{n}^{\mp}(t)^* \mathfrak{C}_{m}^{\mp}(t) e^{|\beta_{-}|^2} \delta^{(2)}(\beta_{-})\Big),
\label{P_explicit}
\eea
where $\beta_{\pm}=\beta \pm \frac{\lambda}{\omega}$.
The above  $P$ distribution incorporating  derivatives of $\delta$ functions is highly singular.
This is a typical behavior observed for nonclassical states.
Due to its highly nonsingular
nature, employing $P(\beta)$  distribution directly towards producing a quantitative measure of  nonclassicality 
is complicated. So other quasi-probability distributions should be considered in this regard. 

\subsection{The Wigner $W$ function}

\begin{figure}
\begin{center}
\subfloat[]{\includegraphics[width=4.5cm,height=4.5cm]{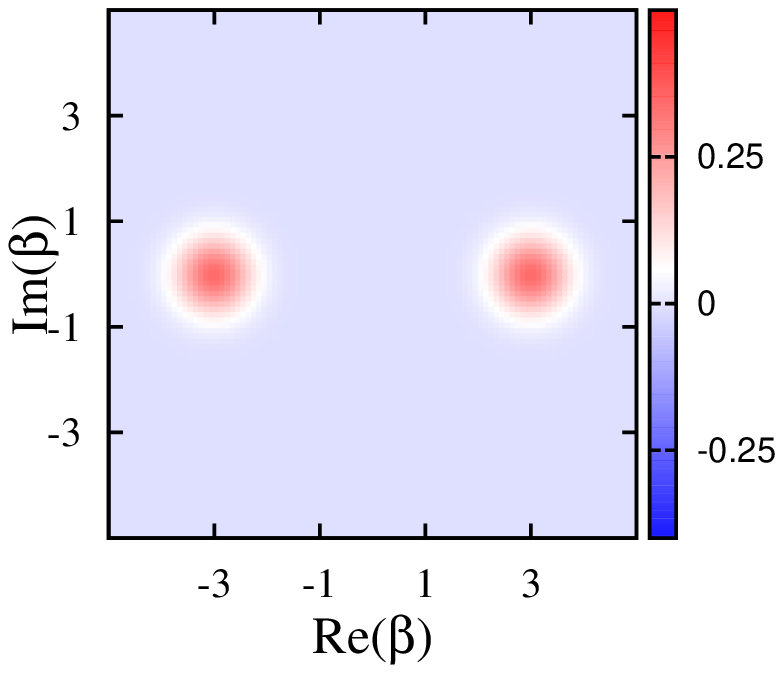}} 
\captionsetup[subfigure]{labelformat=empty}
\subfloat[(b)]{\includegraphics[width=4.5cm,height=4.5cm]{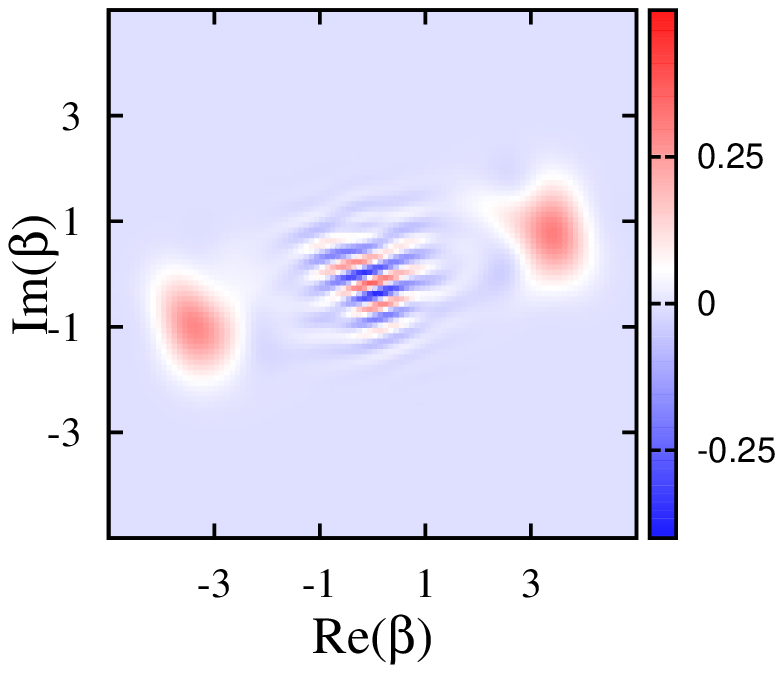}} 
\captionsetup[subfigure]{labelformat=empty}
  \subfloat[(c)]{\includegraphics[width=4.5cm,height=4.5cm]{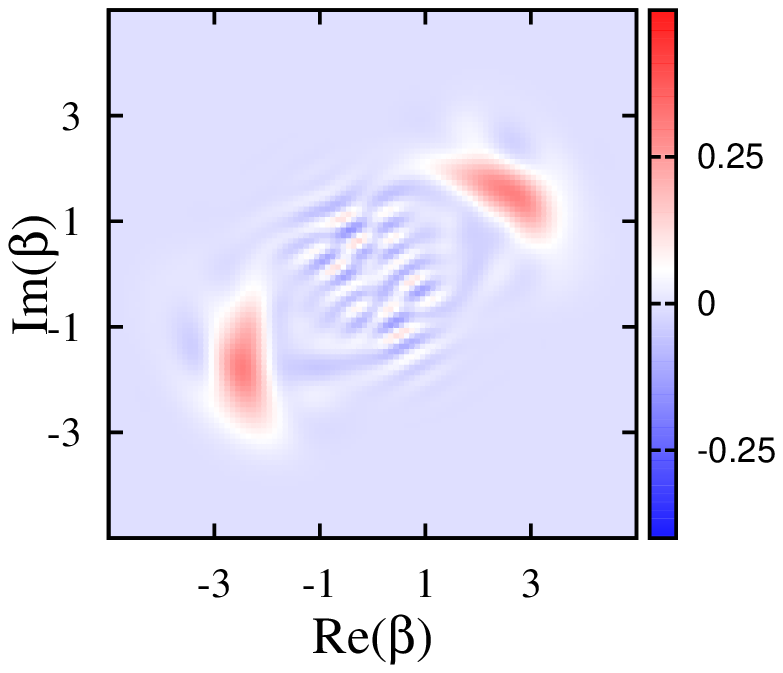}} 
\caption{The density plots of  $W^{(+)}(\beta)$ using (\ref{wigner})  for $ \Delta=0.15\, \omega, \epsilon=0.03 \omega$ and $\alpha=3$ at 
various values of scaled time $\omega t$ (a) $0$, (b) $50$, (c) $100$ 
for the coupling strength  $\lambda=0.3\, \omega$.} 
\label{Wfunc_fig}  
\end{center}
\end{figure}

For an arbitrary density matrix $\rho$ the Wigner quasi-probability distribution is defined [\cite{W1932}]
via the displacement operator as
\beq
W(\beta)=\pi^2 \int \mathrm{Tr}[\rho \mathrm{D}(\gamma)] \exp(\beta \gamma^*-\beta^* \gamma) d^2 \gamma.
\label{W_def}
\eeq
But the evaluation of the Wigner function using the definition (\ref{W_def}) is not always easy. An alternate
series representation of the  distribution $W(\beta)$ in terms of  the diagonal matrix elements in the displaced number states is known
[\cite{CK1993}]:
\beq
W(\beta)=\dfrac{2}{\pi} \sum^{\infty}_{k=0} (-1)^k \langle \beta,k|\rho| \beta, k \rangle,
\qquad | \beta ,k \rangle = \mathrm{D}(\beta) | k \rangle.
\label{Wigner_series}
\eeq
Substituting (\ref{o_density_matrix}) in  (\ref{Wigner_series}) we obtain the time evolution of the Wigner function for the initial quasi-Bell states:
\bea
W^{(\pm)}(\beta)&=&\frac{1}{\pi} \exp(-|\alpha_{+}|^2) \sum_{k=0}^{\infty} (-1)^k \left \lgroup 
 \left | \sum_{m=0}^{\infty} \frac{(\alpha_{+}^*)^m  \mathfrak{C}_{m}^{\pm}(t)^*}{\sqrt{m!}} 
D_{mk}(\beta_{+}) e^{i\omega m t}\right |^2  \right. \nn \\
&& + \left .   \left | \sum_{m=0}^{\infty} (-1)^m \frac{(\alpha_{+}^*)^m  \mathfrak{C}_{m}^{\mp}(t)}{\sqrt{m!}} 
D_{mk}(\beta_{-}) e^{i\omega m t}\right |^2 \right \rgroup.
\label{wigner_inter}
\eea
The component 
$D_{mk}(\beta)$ is evaluated using  the hypergeometric function as
\beq
D_{mk}(\beta) = (-1)^k \exp(-|\beta|^2/2) \frac{(\beta)^m (\beta^*)^k}{\sqrt{m! k!}} {}_2F_0 \left(-m,-k;\underline{\phantom{x}};-\frac{1}{|\beta|^2}\right),
\eeq
where the generalized hypergeometric function ${}_2F_0 (m,n;\underline{\phantom{x}};z)$ reads 
\beq
{}_2F_0 ( m,n;\underline{\phantom{x}};z)=\sum_{r=0}^{\infty} \frac{(m)_r (n)_r}{r!} z^r.
\eeq
The identity
\beq
\sum_{k=0}^{\infty} \frac{(-1)^k x^k}{k!} {}_2F_0 \Big ( -k,-m;\underline{\phantom{x}};-\frac{1}{x} \Big)
 {}_2F_0 \Big  (-k,-n;\underline{\phantom{x}};-\frac{1}{x} \Big)=2^{n+m} e^{-x}  {}_2F_0 \Big ( -m,-n;\underline{\phantom{x}};-\frac{1}{4x} \Big)
\eeq
yields the Wigner function (\ref{wigner_inter}) in the following form:
\bea
W^{(\pm)}(\beta) &=& \!\! \frac{1}{\pi} \exp(-|\alpha_{+}|^2) \!\! \sum_{n,m=0}^{\infty} \! \!
 \frac{(2\alpha_{+})^n   (2\alpha_{+}^*)^m }{n! m!}  \exp(-i\omega (n-m) t)
  \left \lgroup \exp(-2|\beta_{+}|^2)  \right.   \times \qquad \nn \\  
 & \times & (\beta_{+}^*)^n (\beta_{+})^m \mathfrak{C}_{n}^{\pm}(t) \mathfrak{C}_{m}^{\pm}(t)^* 
{}_2F_0 \left( -m,-n;\underline{\phantom{x}};-\frac{1}{4|\beta_{+}|^2} \right)
  + (-1)^{n+m}   \qquad \times  \nn \\
  &  \times & \left. \exp(-2|\beta_{-}|^2) (\beta_{-}^*)^n  
(\beta_{-})^m  
\mathfrak{C}_{n}^{\mp}(t)^* \mathfrak{C}_{m}^{\mp}(t) 
{}_2F_0 \left( -m,-n;\underline{\phantom{x}};-\frac{1}{4|\beta_{-}|^2} \right) \right \rgroup. 
\label{wigner}
\eea
The expression (\ref{wigner}) satisfies the normalization condition:
\beq
 \int W^{(\pm)}(\beta) d^2\beta=1.
\label{W_normalization}
\eeq  
Fig.\ref{Wfunc_fig} shows the Wigner function at three different instances of time. The  distribution for the initial state consists of two Gaussian peaks as it is the superposition of two well-separated coherent states. As time evolves it assumes negative values  demonstrating the 
nonclassical nature of the state.  For a suitably strong qubit-oscillator coupling a large number of interacting modes set in. The quantum interference between these modes give rise to negative values of the $W$-distribution in the zone of the phase space intermediate between the positive peaks.  The volume of the negative domain of the Wigner function on the phase space is considered as 
a quantitative measure of nonclassicality of the density matrix [\cite{KZ2004}]: 
\beq
\Lambda^{(\pm)} = \int |W^{(\pm)}(\beta)| d^2 \beta -1.
\label{negativity_def}
\eeq
A non-zero value of $\Lambda^{(\pm)}$ indicates the characteristic quantum nature of the given state. The time evolution of $\Lambda^{(\pm)}$  for various coupling regime is discussed
in the following subsection.
 \subsection{Husimi Q function }
The Husimi $Q$ function [\cite{H1940}] is defined as the expectation value of  the  density matrix of the oscillator  in an arbitrary coherent state:
\beq
Q(\beta)=\dfrac{1}{\pi} \langle \beta| \rho|\beta\rangle.
\label{Q_def}
\eeq
For the state (\ref{o_density_matrix}) under consideration  the $Q$-function assumes the explicit positive definite form
\bea
Q^{(\pm)}(\beta)=\dfrac{1}{2 \pi} \exp(-|\alpha_{+}|^2) \Big (  \exp(-|\beta_{+}|^2) | \mathfrak{X}^{\pm}|^2 + 
\exp(-|\beta_{-}|^2) | \mathfrak{Y}^{\pm}|^2
 \Big).
\label{Q_factorized}
\eea
The Fourier sums in (\ref{Q_factorized}) are given by
\beq 
\mathfrak{X}^{\pm} = \sum_{n=0}^{\infty} \dfrac{(\alpha_{+} \beta_{+}^*)^{n}}{n!}\mathfrak{C}_{n}^{\pm}(t)\,\exp (-i n \omega t),\quad
\mathfrak{Y}^{\pm} = \sum_{n=0}^{\infty} (-1)^{n} \dfrac{(\alpha_{+} \beta_{-}^*)^{n}}{n!}\mathfrak{C}_{n}^{\mp}(t)^{*}\,\exp (-i n \omega t).
\label{XY_def}
\eeq
 The Husimi distribution (\ref{Q_factorized}) of the reduced density matrix  (\ref{o_density_matrix}) does not have any zero 
on the phase space except at
asymptotically large radial distances. It
 satisfies the normalization condition: $\int Q^{(\pm)}(\beta)\, d^{2}\beta = 1$.
 Adopting the procedure developed in [\cite{IGMS2005}] where the Laguerre functions are approximated by their linear parts in the
regime $\frac{\lambda}{\omega} \lesssim 0.1$ we may approximate the  Fourier sums (\ref{XY_def}) in closed forms as
\bea 
\mathfrak{X}^{(\pm)} &=& \exp ( \Phi_{t}^{\|} )  \cos  \Big( \Phi_{t}^{\perp} - 
\dfrac{\tau}{2} \Big )
\pm i \exp ( - \Phi_{t}^{\|} )  \sin \Big(\Phi_{t}^{\perp}+ \dfrac{\tau}{2} \Big ) \nn \\
&& -i \dfrac{\epsilon}{\widetilde{\Delta}} \exp ( \Phi_{t}^{\|} ) 
\Big \lgroup \sin \Big(\Phi_{t}^{\perp} -
\dfrac{\tau}{2} \Big )+x \, \Phi_t \sin \Big( \Phi_{t}^{\perp} -
\dfrac{(1-x) \tau }{2} \Big ) \Big \rgroup \nn \\ 
&& \mp i \dfrac{\epsilon^2}{2 {\widetilde{\Delta}}^2} \exp ( - \Phi_{t}^{\|} )  \Big \lgroup 
 \sin \Big( \Phi_{t}^{\perp}+\dfrac{\tau}{2} 
\Big ) -2 x \Phi_t \sin \Big ( \Phi_{t}^{\perp}  +\dfrac{(1-x) \tau}{2} \Big )\Big \rgroup,\nn\\
\mathfrak{Y}^{(\pm)} &=& \exp (- \Psi_{t}^{\|}  )  \cos  \Big( \Psi_{t}^{\perp} + \dfrac{\tau}{2} \Big)
\mp i \exp (\Psi_{t}^{\|}  )  \sin \Big(\Psi_{t}^{\perp} - \dfrac{\tau}{2} \Big ) \nn \\
&& -i \dfrac{\epsilon}{\widetilde{\Delta}} \exp ( -\Psi_{t}^{\|}  ) \Big \lgroup \sin \Big( \Psi_{t}^{\perp} +
\dfrac{\tau}{2} \Big )-x \, \Psi_t \sin \Big( \Psi_{t}^{\perp} +
\dfrac{(1-x) \tau}{2} \Big ) \Big \rgroup \nn \\ 
&& \pm i \dfrac{\epsilon^2}{2 {\widetilde{\Delta}}^2} \exp (  \Psi_{t}^{\|}  )  \Big \lgroup 
\sin \Big( \Psi_{t}^{\perp} -\dfrac{\tau}{2} 
\Big ) +2 x \, \Psi_t \sin \Big( \Psi_{t}^{\perp}   -\dfrac{(1-x) \tau}{2} \Big )\Big \rgroup.
\label{Q_linear} 
\eea
In (\ref{Q_linear}) we have used the notations  
$\Phi_t= \alpha_{+}\beta_{+}^* \exp(-i \omega t), \; \Psi_t=\alpha_{+}\beta_{-}^* \exp(-i \omega t),
\Phi_{t}^{\|}=\Phi_t \cos \dfrac{x \tau}{2},\;  
 \Psi_{t}^{\|}=\Psi_t \cos \dfrac{x \tau}{2}, \; \Phi_{t}^{\perp}= \Phi_t \sin \dfrac{x \tau}{2},  $ 
  $ \Psi_{t}^{\perp}=\Psi_t \sin \dfrac{x \tau}{2}$ and $\tau = \widetilde{\Delta}\, t$.

Employing the following well-known interrelations [\cite{C1998}] the Wigner and Husimi functions can also be  directly obtained from the $P$ distribution:
\beq
Q(\beta)=\frac{1}{\pi}\int  e^{-|\beta-\gamma|^2} P(\gamma) d^2 \gamma, \qquad
W(\beta)= \frac{1}{\pi}\int  e^{-2|\beta-\gamma|^2} P(\gamma) d^2 \gamma.
\label{QW_relate_P}
\eeq 
On the other hand, the following property [\cite{C1998}] suggests that the $Q$ function may be considered as  a `coarse-grained' behavior of the  $W$ function:
\beq
Q(\beta)= \frac{1}{\pi}\int  e^{-2|\beta-\gamma|^2} W(\gamma) d^2 \gamma,
\label{Q_W}
\eeq
where the $Q$ function is obtained after a suitable `smearing' of the $W$ function with a positive definite kernel.
The results obtained earlier for the quasi-probability distributions using the direct formulae and the interrelation are found to be consistent. 
To understand the role of negativity of the Wigner distribution we now do a comparative study of the Wehrl entropy [\cite{W1978}]
that is based on the $Q$ function and the recently proposed [\cite{SKD2012}] measure of entropy based on the Wigner $W$ 
function. The Wehrl entropy [\cite{W1978}] defined as 
\beq
S^{(\pm)}_{Q} = - \int  Q^{(\pm)}(\beta)\, \log Q^{(\pm)}(\beta) d^{2}\beta
\label{Wehrl} 
\eeq
acts as an information-theoretic measure describing  the delocalization of the oscillator on the phase space. It is also of interest to study the quantum entropy based on the modulus of the Wigner distribution $|W(\beta)|$ [\cite{SKD2012}] which is a non-negative quantity:
\beq
S^{(\pm)}_{W}=-\int |W^{(\pm)}(\beta)| \log |W^{(\pm)}(\beta)| d^2 \beta.
\label{S_W}
\eeq
\begin{figure}
 \begin{center}
\subfloat[]{\includegraphics[scale=1.25]{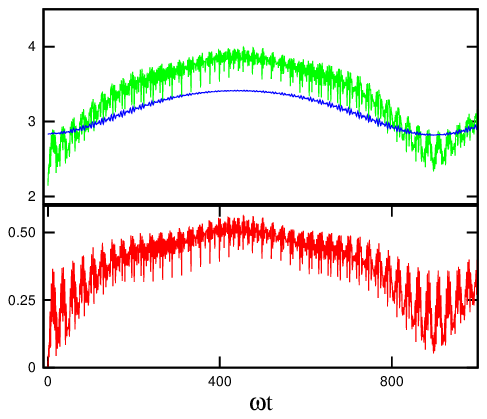}}
\subfloat[]{\includegraphics[scale=1.25]{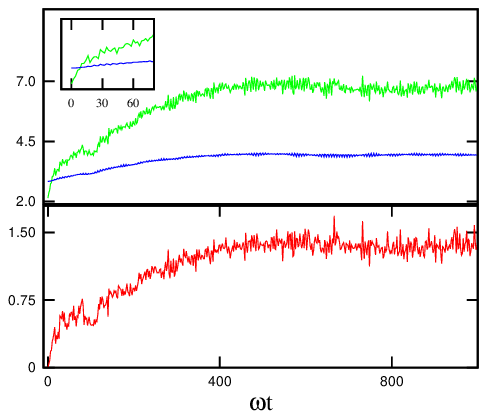}}
 \end{center}
\vspace{-1.5cm}
 \caption{The time evolutions of  negativity $\Lambda^{(+)}$ (red), Wigner entropy $S_W^{(+)}$ (green) and Wehrl entropy $S_Q^{(+)}$  obtained for the values of
$ \Delta=0.15 \,\omega, \epsilon=0.03 \,\omega$ at
(a) $\alpha=2$,   $\lambda$=0.08\,$\omega$ and (b) $\alpha=3$, $\lambda$= 0.3\,$\omega$.}
\vspace{-0.25cm}
\label{comp_fig}
\end{figure}
\par
It is evident from the Figs. \ref{comp_fig} (a) and (b) that the time evolutions of the Wigner entropy (\ref{S_W}) and  negativity parameter (\ref{negativity_def}) have close kinship with each other. An increasing $S_W^{(+)}$ reciprocates increment in $\Lambda^{(+)}$ and \textit{vice versa}.
In this sense the Wigner entropy  $S_W^{(+)}$ reveals the extent of nonclassicality of a quantum density matrix. We distinguish between two possible scenarios depending upon the qubit-oscillator coupling strength. {\sf (i)} In the strong coupling regime $(\lambda/\omega \lesssim 0.1)$ we observe (Fig. \ref{comp_fig} (a)) that a periodic structure in the phase space distribution develop, and,
consequently, the physical variables such as the  Wigner entropy $S_W^{(+)}$, negativity $\Lambda^{(+)}$ and Wehrl entropy $S_Q^{(+)}$ follow similar periodic patterns that may be identified 
with the revival and collapse of the qubit density matrix elements. For a dominant value of  $\Lambda^{(+)}$ when quantum interference effects are overwhelming, we, expectedly, find 
$S_W^{(+)} > S_Q^{(+)}$. On the other hand, for a low $\Lambda^{(+)}$ region the inequality is reversed: $S_W^{(+)} < S_Q^{(+)}$.  As the $Q$-function is obtained from the $W$-function 
(\ref{QW_relate_P}) after suitable averaging with a positive definite kernel, the Wehrl entropy $S_Q^{(+)}$ would be larger in a low negativity $\Lambda^{(+)}$ regime [\cite{MAN2000}]. {\sf (ii)} In the ultra strong coupling regime  $(\lambda/\omega \gg 0.1)$ all 
Fourier modes for the qubit-oscillator
 interaction are excited and a fully randomized interference pattern very quickly evolves. The randomized interferences of a large number of Fourier modes lead to quasi-stationary values of the phase space observables (Fig. \ref{comp_fig} (b)). Moreover, these interferences  necessarily develop areas on the phase space with negative values of $W$-distribution. The average value of the negativity parameter $\Lambda^{(+)}$ increases with increase in the coupling strength. To keep, however,  the normalization sum rule (\ref{W_normalization}) intact, suitable increment in the  magnitude of the $W$ function takes place leading to increased value of the entropy $S_{W}$. In the quasi-stationary state the negativity $\Lambda^{(+)}$ is statistically preserved. The quasi-stabilization of $\Lambda^{(+)}$ occurs after a suitable decoherence time. Consequently, we observe that except for a brief initial period the Wigner entropy $S_W^{(+)}$ is consistently more than the Wehrl entropy $S_Q^{(+)}$, even though the positive definite $Q$  function may viewed as the smeared form of the $W$ distribution. 

\section{Conclusion}
Phase space dynamics of the strongly coupled qubit-oscillator system is studied under adiabatic approximation. Using this approximation
the oscillator density matrix have been written in terms of the displaced number states. This density matrix is employed to calculate the 
quasi-probability distributions such as Glauber-Sudarshan $P$ function, Wigner $W$ function  and the Husimi $Q$ function. The 
negative values of the Wigner distribution acts as an witness of the nonclassicality of the state.   The time evolution of this nonclassicality parameter of the state is obtained. In the ultra strong coupling regime the negativity assumes, after a suitable decoherence time,  a quasi-stationary value. The quantum entropy based on Wigner function  
 exhibits qualitatively similar behavior as that of the nonclassicality. The value on the nonclassicality measure serves as the key to understand the relative values of the entropies based on Wigner and Husimi distributions.

\section*{Acknowledgement}
One of us (BVJ)  acknowledges the support from UGC (India) under the Maulana Azad National Fellowship  scheme.


\begin{thebibliography}{99}
\bibitem{JC1963} E.T. Jaynes, F.W. Cummings, Proc. IEEE {\bf 51}, 89 (1963).

\bibitem{ABS2002} A.D. Armour, M.P. Blencowe, K.C. Schwab, Phys. Rev. Lett. {\bf 88}, 148301 
(2002).
 
\bibitem{ALT2009} A.A. Anappara, S.D. Liberato, A. Tredicucci, C. Ciuti, G. Biasiol, L. Sorba, F. Beltram,
Phys. Rev. B {\bf 79}, 201303 (2009).
  
\bibitem{ND2010} T. Niemczyk, F. Deppe,  H. Huebl, E.P. Menzel, F. Hocke, M.J. Schwarz,
J.J. Garcia-Ripoll, D. Zueco, T. H\"{u}mmer, E. Solano,  A. Marx, R. Gross, Nature Physics {\bf 6}, 772 (2010).

\bibitem {IGMS2005} E.K. Irish, J. Gea-Banacloche, J. Martin, K.C. Schwab, Phys. Rev. B {\bf 72}, 195410 (2005).

\bibitem{AN2010} S. Ashhab, F. Nori, Phys. Rev. A {\bf 81}, 042311 (2010).

\bibitem{W1932} E. P. Wigner, Phys. Rev. {\bf 40}, 749 (1932).

\bibitem{KZ2004} A. Kenfack and K. Zyczkowski, J. Opt. B {\bf 6}, 396 (2004).

\bibitem{H1940} K. Husimi, Proc. Phys. Math. Soc. Jpn. {\bf 22}, 264 (1940).

\bibitem{MAN2000}G. Manfredi and M.R. Feix, Phys. Rev. E {\bf 62}, 4665 (2000).

\bibitem{SKD2012} P. Sadeghi, S. Khademi, A.H. Darooneh
Phys. Rev. A {\bf 86}, 012119 (2012).

\bibitem{G1963} R.J. Glauber, Phys. Rev. {\bf 131}, 2766 (1963).

\bibitem{S1963} E.C.G. Sudarshan, Phys. Rev. Lett. {\bf 10}, 277 (1963).

\bibitem{W1978} A. Wehrl, Rev. Mod. Phys. {\bf 50}, 221 (1978).

\bibitem{R2013} R. Chakrabarti, B.V. Jenisha, {\textit{Quasi-Bell states in a strongly coupled qubit-oscillator system and their delocalization in phase space}, arXiv:1302.2771v3 (2014).}

\bibitem{CK1993} H. Moya Cessa, P.L. Knight,
Phys. Rev. A {\bf 48}, 2479 (1993).

\bibitem{C1998} H.J. Carmichael, {\textit{Statistical Methods in Quantum Optics I}}, Springer-Verlag, Berlin (1998).

\end{thebibliography}
\end{document}